\documentclass[traditabstract]{aachanged}
\usepackage[english]{babel}
\usepackage{longtable}
\usepackage{graphicx}	
\usepackage{epsfig}
\usepackage{epsf}
\usepackage{amsmath}	
\usepackage{amssymb}	
\usepackage{xcolor,soul}
\usepackage{subcaption}
\usepackage{soul}
\usepackage[normalem]{ulem}

\newcommand{\CI}[0]{{}C\,{\sc i}\,}

\begin{document}

\title{\bf Estimation of the CMB temperature \\ from atomic C\,{\sc i} and molecular CO lines \\ in the interstellar medium of early galaxies}
\titlerunning{Measurements of the CMB temperature}
\author{V.V. Klimenko$^{1}$\thanks{E-mail: slava.klimenko@gmail.com}, A.V. Ivanchik$^1$, P. Petitjean$^2$, P. Noterdaeme$^2$, R. Srianand$^3$}
\authorrunning{Klimenko et al.}
\date{Received  October 27, 2020; revised October 27, 2020; accepted October 27, 2020}

\institute{\it{$^{1}$Ioffe Institute, Russian Academy of Sciences, ul. Politekhnicheskaya 26,
St. Petersburg, 194021 Russia}\\
\it{$^{2}$Institut d’Astrophysique de Paris, CNRS-UPMC, UMR7095, 98bis Bd Arago, 75014 Paris, France}\\
\it{$^3$Inter-University Centre for Astronomy and Astrophysics, Post Bag 4, Ganeshkhind, 411 007, Pune, India}}

\abstract{The linear increase of the cosmic microwave background (CMB) temperature with cosmological redshift, $T_{\rm CMB} = T_0(1 + z)$, is a prediction of the standard cosmological $\Lambda$CDM model. There are currently two methods to measure this dependence at redshift $z>0$, and that is equally important to estimate the CMB temperature $T_0$ at the present epoch $z=0$. The first method is based on the Sunyaev–Zeldovich (SZ) effect for a galaxy cluster. However, this method is limited to redshifts $z\le1$ and only the deviations from the standard relation can be measured with it. The second method is based on the analysis of the populations of atomic and molecular energy levels observed in the absorption spectra of quasars. This method allows $T_{\rm CMB}(z)$ to be measured directly. We present new estimates of $T_{\rm CMB}(z_i)$ in the redshift range $1.7\le z_i \le3.3$ based on the analysis of excitation of the CO rotational levels and C\,{\sc i} fine-structure levels in 15 absorption systems. We take into account collisional excitation of CO and C\,{\sc i} with hydrogen atoms and H$_2$ and radiative pumping of C\,{\sc i} by the interstellar ultraviolet radiation. Applying this corrections leads to a systematic decrease in the previously obtained estimates of $T_{\rm CMB}(z_i)$ (for some systems the magnitude of the effect is $\sim$10\%). Combining our measurements with the measurements of $T_{\rm CMB}(z)$ in galaxy clusters we have obtained a constraint on the parameter $\beta=+0.010\pm0.013$, which characterizes the deviation of the CMB temperature from the standard relation, $T_{\rm CMB} = T_0(1 + z)^{1-\beta}$, and an independent estimate of the CMB temperature at the present epoch, $T_0 = 2.719\pm0.009$\,K, which agrees well with the estimate from orbital measurements, $T_0 = 2.7255\pm0.0006$\,K. This independent estimate is very important because it was obtained using cosmological data, in contrast to satellite measurements, which are obtained ``here'' and ``now''.}

\keywords{cosmology, early Universe, interstellar medium, quasar spectra.}

\maketitle

\section{Introduction}

Having been born in the first instants of the Big Bang, the cosmic microwave background (CMB) plays a crucial role in the post-inflation expansion dynamics of the Universe at the early stage of its evolution. Within fractions of a second after the completion of the inflationary stage, the Universe passes  to the radiation-dominated stage, during which its expansion rate is determined by relativistic matter (mostly photons and neutrinos) whose energy density is much greater than the energy of all other forms of matter (baryonic matter, dark matter, and dark energy). Only approximately 50 000 years after the Big Bang (see, e.g., Gorbunov and Rubakov 2016) does non-relativistic matter (dark matter and baryons) begin to dominate in the energy density and to change the expansion rate of the Universe, transferring it to the stage of dominance of non-relativistic matter.

Apart from the expansion dynamics of the Universe, the CMB plays a crucial role in two physical processes that are cosmological markers in the evolution of the Universe. These are primordial nucleosynthesis, which began approximately 180\,s (3 min) after the Big Bang and lasted slightly more than three hours, and primordial hydrogen–helium plasma recombination, which proceeded 380\,000 years after the Big Bang. Both these processes proceed at distinctly different cosmological epochs,
nuclear physics in the first case and atomic physics in the second case are used for their description, for each of them there are observational data that provide independent estimates for the key cosmological parameters. 

At present, three key properties of the CMB attract particular attention, because their study gives important cosmological information. These include (i) the direct Planck blackbody spectrum (Fixsen et al. 1996) and its distortions, (ii) the angular CMB temperature anisotropy (Planck Collaboration 2020), and (iii) the CMB polarization whose studies can lead to the detection of primordial gravitational waves. Yet another important prediction of the standard cosmological model is the dependence of the CMB temperature on cosmological redshift $z$: 
\begin{equation}
\label{eq:T_z}
    T_{\rm CMB}=T_0(1+z)
\end{equation}
where $T_0 = 2.7255 \pm 0.0006$\,K is the current CMB temperature (Fixsen 2009). In such a form this relation has held from the epoch of primordial nucleosynthesis until now. Its validity has been tested with numerous observations, consistent with the primordial nucleosynthesis and primordial recombination; however, as yet there are no direct observations confirming this relation for these epochs. Furthermore, there are various alternative cosmological models which predict a deviation of the redshift dependence of the CMB temperature from the standard relation (see, e.g., Freese et al. 1987; Matyjasek 1995).

In this paper we focus our attention on the possibility of testing the standard law of evolution of the CMB temperature and on the methods for an independent estimation of the current CMB temperature $T_0$.

\section{CMB temperature measurements}

Observational indication of the CMB were detected long before the prediction of this phenomenon itself by Gamow and Alpher and Herman (1946, 1948). While studying the CN molecules in the interstellar medium (ISM) of our Galaxy, McKellar (1940) found that not all of the CN molecules are in the ground energy state (as was originally assumed for a diffuse cold medium) and some of the molecules are in an excited state as if they were excited by thermal radiation with a temperature $T \sim 3$\,K. Only much later, when Penzias and Wilson (1965) discovered the CMB radiation, did it become clear that this radiation permeates the entire space and excites molecules. Roth and Meyer (1993) performed more accurate measurements of the excitation of CN molecules in the ISM of our Galaxy and obtained an estimate, $T_0 = 2.729^{+0.023}_{-0.031}$\,K, that agrees well with the currently most accurate estimate from the COBE/FIRAS and WMAP space experiments: $T_0 = 2.7255 \pm 0.0006$\,K (Fixsen 2009). However, these CMB temperature measurements were performed “here” and “now”, i.e., in the Solar System and at the present epoch ($z = 0$). There are currently two methods to measure the CMB temperature at redshift $z>0$. 

The first method is based on the analysis of the Sunyaev–Zeldovich (SZ) effect for galaxy clusters (Zeldovich and Sunyaev 1969) and was proposed 10 years after the prediction of the SZ effect by Fabbri et al. (1978) and Rephaeli (1980). Despite the fact that the galaxy clusters are at different cosmological redshifts $z_i$, the distortion of the CMB spectrum caused by the interaction of CMB photons with a hot electron gas is $z$-independent within the standard cosmological model. This is because the dependence of both the Planck spectrum and its distortion caused by the SZ effect on frequency $\nu$ is defined via the dimensionless parameter $x$,
\begin{equation}
        x=\frac{h\nu_i}{kT_i} =\frac{h\nu_0(1+z_i)}{kT_0(1+z_i)}=\frac{h\nu_0}{kT_0},
\end{equation}
which is $z$-independent within the standard cosmological  model, where the change in photon frequency and CMB temperature is proportional to the same factor $(1 + z)$. However, within alternative cosmological scenarios the SZ distortions of the Planck spectrum can become $z$-dependent. The deviations of the evolution of the CMB temperature  from the standard relation (1) can be parameterized in the form proposed by Lima et al. (2000):
\begin{equation}
     T(z)=T_0(1+z)^{1-\beta},
\end{equation}
while the redshift dependence of the frequency, being a more general property defined by General Relativity, $\nu(z) = \nu_0(1 + z)$ does not changed. This leads to the z dependence of the spectral shape of the SZ distortion due to the $x(z)$ dependence:
\begin{equation}
     x=\frac{h\nu_0(1+z_i)}{kT_0(1+z_i)^{1-\beta}}=\frac{h\nu_0}{kT_0}(1+z)^{\beta}.
\end{equation}
This underlies the method of searching for possible deviations from the standard law (1). Moreover, within both the standard and alternative theory the SZ effect allows (as was pointed out by Fabbri et al. 1978 and Rephaeli 1980) the current CMB temperature $T_0$ to be independently estimated owing to the spectral peculiarities of the SZ distortion. In particular, there exists a critical value of the dimensionless parameter $x_0 = 3.830$ for which the distortion of the Planck spectrum is zero. Thus, if we measure the wavelength $\lambda_0$ at which the distortion of Planck spectrum towards a galaxy cluster is zero, then we can determine the $T_0=hc/(3830k\lambda_0)$ to be also measured (Fabbri et al. 1978).  

In recent years, in view of the increase in observational statistics, many studies have been performed to test the standard evolution of the CMB temperature (see, e.g., Battistelli et al. 2002; Luzzi et al. 2009; Hurier et al. 2014; Luzzi et al. 2015; Avgoustidis et al. 2016). In our paper we will also use data on the SZ effect.

The second method is based on the analysis of the excitation of atomic and molecular energy levels observed in the absorption spectra of quasars. It was proposed even earlier than the first method by Bahcall and Wolf (1968). This method allows $T_{\rm CMB}(z)$ to be measured directly (see, e.g., Srianand et al. 2000). The most convenient elements in this method are the fine-structure energy levels of atomic C\,{\sc i} and the rotational levels of CO molecules (Silva and Viegas 2002; Srianand et al. 2008; Noterdaeme et al. 2011).

Each of the described methods has its advantages and disadvantages. The $T_{\rm CMB}(z)$ measurements in galaxy clusters have high statistics of the number of systems, but they are limited to the redshift range $z \le 1$, for which the deviation of $T_{\rm CMB}(z)$ from the standard relation is small. In contrast, the $T_{\rm CMB}(z)$ measurements using the analysis of atomic and molecular absorption lines correspond to redshifts $z\sim2–3$, at which the possible deviation from the standard relation can be more significant. However, the probability to detect molecular systems in the spectra of quasars is rather low, $\sim4$\% (Balashev and Noterdaeme 2018) due to the molecular clouds being compact (for example, in comparison with atomic absorption clouds). The second, this method requires quasar spectra obtained with high signal to noise ratio and high spectral resolution. At present, CO absorption lines with $z\sim2$ have been detected in six quasar spectra (Noterdaeme et al. 2011, 2018) and about 20 C\,{\sc i} absorption systems at redshift $2\sim3$ have been detected in quasar spectra with high resolution. The main source of the systematic effects related to the uncertainty of the physical conditions in absorption systems, which often cannot be measured well. Therefore, only upper limits on $T_{\rm CMB}(z)$ have been set for most C\,{\sc i} absorption systems. Klimenko and Balashev (2020) showed that a consistent analysis of the excitation of C\,{\sc i} fine-structure levels and molecular hydrogen (H$_2$) rotational levels allows the physical conditions (UV background intensity, gas number density, the kinetic temperature) in absorption systems to be determined more reliably.

In conclusion of this section, note once again that both these methods allow to independently estimate the current CMB temperature $T_0$ that was proposed in early papers and which is occasionally overlooked later. This can be done by extrapolating the dependence T(z) to zero redshift ($z = 0$), i.e., by considering $T_{\rm 0}$ as a free parameter when fitting the data with the law $T({\rm z}) = T_{\rm 0}(1 + z)^{1-\beta}$ (for $\beta = 0$ and $\beta\neq0$). An increase in statistics on galaxy clusters and absorption systems will potentially allow $T_0$ to be determined with an accuracy comparable to or even better than what is done “here” and “now”.

In this paper we present our new measurements of $T_{\rm CMB}(z)$ in C\,{\sc i} and CO absorption systems with $z > 1.7$ made by applying the correction for collisional (for C\,{\sc i} and CO) and radiative pumping by interstellar UV radiation (for C\,{\sc i}) and an independent estimate of the current CMB temperature $T_0$.

\section{Data}

To measure the CMB temperature, we chose 15 damped Lyman-$\alpha$ absorption systems (DLAs) with high redshifts $z > 1.7$ that have high molecular hydrogen column densities $\log N({\rm H_2}) > 18$ (where $N$ is column density measured in ${\rm cm^{-2}}$) and associated C\,{\sc i} absorption lines. CO absorption lines were detected in 6 DLAs (see Noterdaeme et al. 2018b). A list of C\,{\sc i} and CO absorption systems is presented below in Table\,\ref{tab:ci} and Table\,\ref{tab:co}. The observations were performed with the high-resolution spectrographs UVES (Dekker et al. 2000) at the ESO VLT-UT2 telescope in Chili and HIRES (Vogt et al. 1994) at the Keck telescope in Hawaii. The parameters of the observations and the primary analysis of the spectra are described in Noterdaeme et al. (2018) (J\,0000$+$0048), Klimenko et al. (2015) (B\,0528$-$2508), Balashev et al. (2010) (for J\,0812$+$3208, Keck), Guimarres et al. (2012) (J\,0816$+$1446), Balashev et al. (2017) (J\,0843$+$0221), Noterdaeme et al. (2011) (J\,0857$+$1855, J\,1047$+$2057, J\,1705$+$3543), Balashev et al. (2011) (J\,1232$+$0815), Noterdaeme et al. (2010) (J\,1237$+$1047), Srianand et al. (2008) (J\,1439$+$1118), Ledoux et al. (2003) (B\,1444$+$0126), Ranjan et al. (2018) (J\,1513$+$0352), Jorgenson et al. (2010) (J\,2100$-$0641, Keck), and Noterdaeme et al. (2015) (J\,2140$-$0321). For most DLAs we use the measured populations of H$_2$ and CO rotational levels and C\,{\sc i} fine-structure levels. In addition, we determined the parameters of the C\,{\sc i} and CO absorption systems in the DLAs J\,0857$+$1855, J\,1047$+$2057, and J\,1705$+$3543. The results of our measurements are presented in Table\,\ref{table1}. To analyse absorption systems we use the standard procedure of comparing the observed and synthetic spectra described in our previous papers (see, e.g., Balashev et al. 2017; Balashev et al. 2019).

\begin{table*}
\begin{center}
\caption{A list of the high-redshift H$_2$/C\,{\sc i} absorption systems.} 
\label{tab:ci}
\begin{tabular}{l|l|l|l|l|l|l}
\hline
QSO & $z_{\rm abs}$ &   $T_{\rm 02}({\rm H_2})$ & $\log n_{\rm H}$ & $\log I_{\rm UV}$ & $T_{\rm CMB}(\text{C\,{\sc i}}\!)$ & $f_{\rm CMB}^{\rm CI}({\rm J=1})$\\
 &  &  K & $\rm{cm}^{-3}$ & Mathis field & K & \\
\hline 
J\,0000$+$0048 & 2.5255  & $97^{+4}_{-4}$ & $1.49^{+0.25}_{-0.70}$ & $<1$& $11.1^{+1.5}_{-6.6}$ & $0.90^{+0.40}_{-0.30}$ \\ 
B\,0528$-$2505 & 2.8111 & $166^{+8}_{-8}$ & $2.49^{+0.07}_{-0.11}$ & $1.15^{+0.15}_{-0.15}$& none & $0.24\pm0.09$ \\ 
J\,0812$+$3208 & 2.6264  & $52^{+3}_{-3}$ &  $2.55^{+0.16}_{-0.18}$ & $0.04^{+0.21}_{-0.23}$& none  & $0.45^{+0.30}_{-0.20} $  \\ 
               & 2.6263 &    $110^{+5}_{-5}$ & $1.79^{+0.24}_{-0.49}$ & $-0.13^{+0.26}_{-0.30}$ & $10.8^{+1.4}_{-3.3}$ & $0.66\pm0.07$   \\ 
J\,0816$+$1446 & 3.2874  &  $80^{+6}_{-5}$ & $1.77^{+0.45}_{-0.80}$ & $-0.08^{+0.39}_{-0.50}$& $15.2^{+1.0}_{-4.2}$ & $0.60\pm0.04$ \\
J\,0843$+$0221 & 2.7866 &   $123^{+8}_{-8}$ & $1.94^{+0.12}_{-0.10}$ & $1.83^{+0.12}_{-0.13}$& $<16$ & $0.21\pm0.02$ \\ 
J\,1232$+$0815 & 2.3377 &   $64^{+4}_{-4}$ & $2.03^{+0.17}_{-0.18}$ & $-0.13^{+0.40}_{-0.37}$ & $<9.4$ & $0.45\pm0.03$\\ 
J\,1237$+$0647 & 2.6896 &   $178^{+102}_{-50}$ & $1.19^{+0.18}_{-0.17}$ & $0.87^{+0.18}_{-0.15}$ & $<13.8$ &$0.47\pm0.04$ \\ 
J\,1439$+$1118 & 2.4184 &  $117^{+15}_{-17}$ & $0.98^{+0.20}_{-0.25}$ & $0.68^{+0.20}_{-0.24}$ & $<13.7$ & $0.41\pm0.04$ \\
B\,1444$+$0126 & 2.0870 & $172^{+32}_{-23}$ & $2.16^{+0.27}_{-0.26}$ & $0.59^{+0.25}_{-0.25}$ & $<10.5$ & $0.42^{+0.20}_{-0.13}$\\ 
J\,1513$+$0352 & 2.4636  & $89^{+4}_{-4}$ & $1.95^{+0.16}_{-0.36}$ & $0.40^{+0.40}_{-0.69}$ & $<12$ & $0.38\pm0.08$ \\
J\,2100$-$0641 & 3.0915 & $84^{+3}_{-3}$ & $2.02^{+0.15}_{-0.93}$ & $<-0.10$ & $12.9^{+3.3}_{-4.5}$ & $0.63\pm0.13$ \\ 
J\,2140$-$0321 & 2.3399 & $83^{+5}_{-4}$ & $2.93^{+0.23}_{-0.18}$ & $1.54^{+0.20}_{-0.21}$ & none & $0.10\pm0.02$ \\ 
\end{tabular}
\end{center}
\end{table*} 

\begin{table*}
\begin{center}
\caption{A list of high-redshift CO absorption systems.}
\label{tab:co}
\begin{tabular}{c|c|c|c|c|c|c}
\hline
Name &  z & $T^{\rm{exc}}_{\rm 0-J}({\rm CO})$ & $T_{\rm CMB}(z)$ & $\log n_{\rm H}$ & $T_{\rm 02}({\rm H_2})$ & $T_{\rm CMB}({\rm CO})$  \\
 &  & K & K & $\rm{cm}^{-3}$ & K & K  \\
\hline 
J\,0000$+$0048 & $2.5244$ & $9.85^{+0.7}_{-0.6}$ & 9.6 & $1.31^{+0.24}_{-0.42}$ & $97^{+4}_{-4}$ & $9.8^{+0.7}_{-0.6}$\\
J\,0857$+$1855 & $1.7294$ & $8.9^{+1.5}_{-1.2}$ & 7.4 & $2.31^{+0.70}_{-0.20}$ & 100 & $7.9^{+1.7}_{-1.4}$ \\ 
J\,1047$+$2057 & $1.7738$ & $6.9^{+0.7}_{-0.7}$ & 7.5& $<2.5$ & 100 &  $6.6^{+1.2}_{-1.1}$ \\
J\,1237$+$0647  & $2.6896$ & $10.5^{+0.8}_{-0.6}$ & 10.1 & $1.27^{+0.14}_{-0.10}$ & $178^{+102}_{-57}$ & $10.4^{+0.8}_{-0.7}$ \\
J\,1439$+$1118 & $2.4184$ & $9.1^{+0.9}_{-0.7}$ & 9.3 & $0.90^{+0.15}_{-0.18}$ & $107^{+33}_{-20}$  & $9.04^{+0.8}_{-0.7}$ \\
J\,1705$+$3543 & $2.0377$ & $9.1^{+1.8}_{-1.4}$ & 8.3 & $2.21^{+0.17}_{-0.68}$ & 100 & $8.6^{+1.9}_{-1.4}$ \\  
\end{tabular}
\end{center}
\end{table*}

\begin{table}
\begin{center}
\caption{Measurements of column densities of C\,{\sc i} fine-
structure levels and CO rotational levels in absorption systems in the spectra of the quasars J\,0857$+$1855, J\,1047$+$2057, and J\,1705$+$3543.}
\label{table1}
\begin{tabular}{c|c|c|c}
Parameter & J\,0857$+$1855 & J\,1047$+$2057  & J\,1705$+$3543\\
\hline 
z & 1.7293 & 1.7738& 2.0377\\ 
\hline
C\,{\sc i}(J=0) & $13.90^{+0.10}_{-0.10}$ & & $14.57^{+0.06}_{-0.06}$\\
C\,{\sc i}(J=1) & $13.67^{+0.08}_{-0.08}$ & & $14.49^{+0.05}_{-0.05}$ \\
C\,{\sc i}(J=2) & $13.23^{+0.07}_{-0.07}$ & &  $13.98^{+0.05}_{-0.05}$\\
 & & & \\
CO(J=0) & $13.08^{+0.08}_{-0.08}$ &$14.53^{+0.21}_{-0.21}$ & $13.46^{+0.06}_{-0.06}$\\
CO(J=1) & $13.01^{+0.08}_{-0.08}$ &$13.63^{+0.18}_{-0.18}$ & $13.75^{+0.09}_{-0.09}$\\
CO(J=2) & $12.91^{+0.08}_{-0.08}$ &$14.19^{+0.10}_{-0.09}$ & $13.37^{+0.10}_{-0.10}$\\
CO(J=3) & $12.30^{+0.22}_{-0.54}$ &$13.23^{+0.11}_{-0.11}$ & $13.36^{+0.16}_{-0.16}$\\
\end{tabular}
\end{center}
\end{table}

\section{Physical conditions in the ISM}
\label{sect:h2}
Neutral carbon and molecular hydrogen are known to be tracers of cold gas in the diffuse phase of the ISM (see, e.g., Srianand et al. 2005; Balashev et al. 2019). Observations show that C\,{\sc i} was detected only in the absorption systems where H$_2$ is present, i.e., the C\,{\sc ii}/C\,{\sc i} transition occurs in the region where hydrogen is already predominantly in the molecular phase. Physically, this can be caused by the absorption of C\,{\sc i}-ionizing photons with energies $<13.6$\,eV (the C\,{\sc i} ionization potential is $11.26$\,eV) by H$_2$ molecules in Lyman and Werner transition lines and an enhanced gas number density in cold gas, which is also favourable condition for formation of H$_2$. We assume that H$_2$ and C\,{\sc i} are spatially linked, and the populations of their energy levels correspond to the same physical conditions in the ISM. 

\begin{figure}[h]
\begin{center}
        \includegraphics[width=0.5\textwidth]{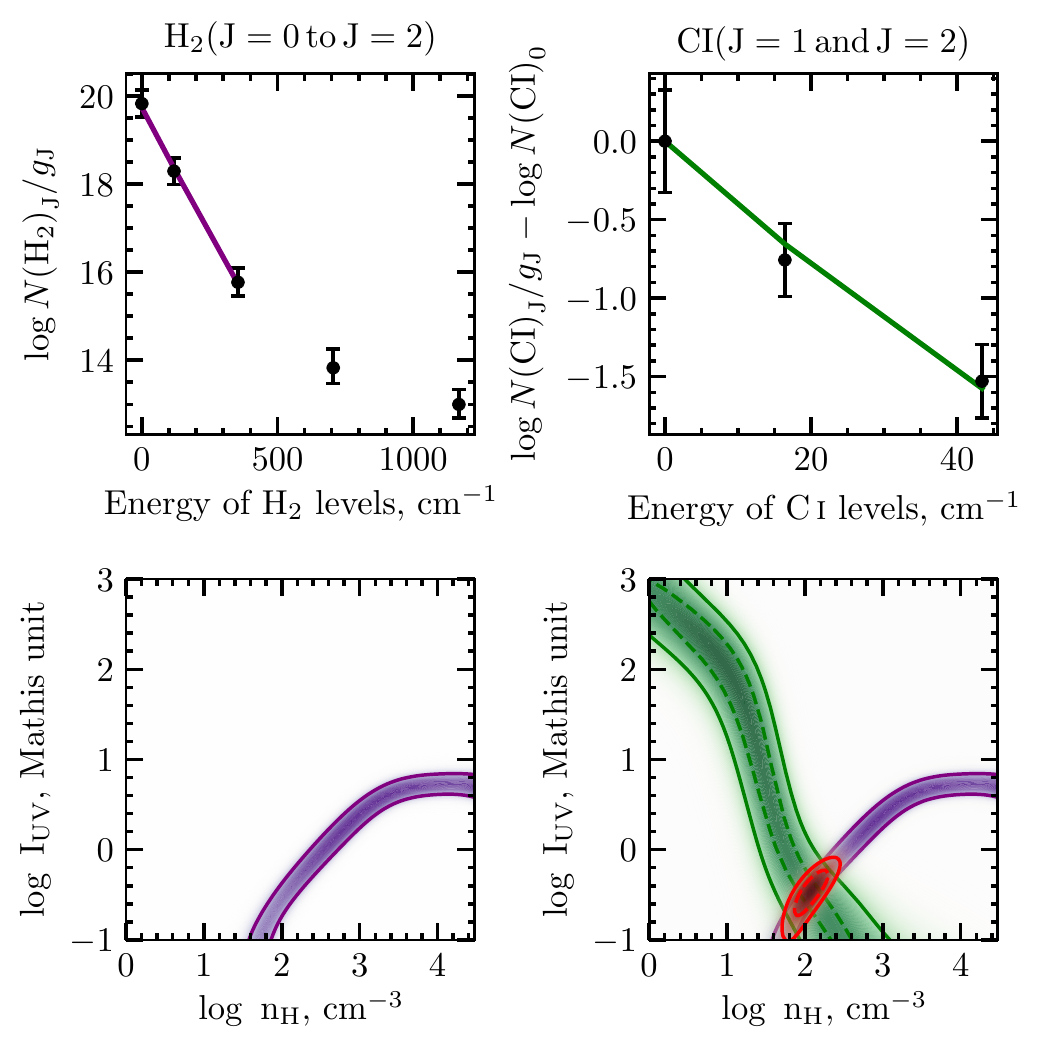}
        \caption{\rm 
        Illustration of the method for determining the physical conditions in an H$_2$ molecular cloud with $z = 2.626443$ in the spectrum of QSO\,J\,0812$+$3208 based on the analysis of the populations of molecular H$_2$ rotational levels and atomic C\,{\sc i} fine-structure levels. On the upper panels the circles indicate the observed H$_2$ (left panel) and C\,{\sc i} (right panel) level populations. The violet and green lines indicate the model level populations for H$_2$ and C\,{\sc i}, respectively. On the lower panels the violet and green contours indicate the constraints on the UV background intensity and the gas number density obtained by analyzing the H$_2$ and C\,{\sc i} level populations. The red contour on the lower right panel indicates the final constraint on the physical conditions using a joint analysis of the H$_2$ and C\,{\sc i} level populations.
        }
        \label{fig:pdr}
\end{center}
\end{figure}

For saturated H$_2$ absorption systems, with $\log N({\rm H_2})>18$, the levels of J=0,1,2 are predominantly thermalized and their excitation is typically close to the thermal temperature (Le Petit et al. 2006), which is set by the thermal balance, itself being a function of the density and ultraviolet (UV) field. Klimenko and Balashev (2020) have shown that an analysis of excitation of lower H$_2$ rotational levels with the PDR Meudon code (Le Petit et al. 2006) allows to constrain a combination of the number density ($n_{\rm H}$) and intensity of UV radiation ($I_{\rm UV}$). In the reasonable range of the number densities corresponded to the cold diffuse medium ($\log n_{\rm H}\sim 2-3$) this translates in almost linear dependence between $I_{\rm UV}$ and $n_{\rm H}$. 
An example of analysis of H$_2$ in the absorption system at $z=2.626$ towards the quasar J\,0812$+$3208 is shown on left panels in Fig.\ref{fig:pdr}. 
An analysis of excitation of C\,{\sc i} levels gives quite wide degenerate region in $I_{\rm UV}-n_{\rm H}$  plane, but it typically is nearly orthogonal to the region constrained using H$_2$ (see right panels in Fig.\,\ref{fig:pdr} and also Balashev et al. 2019). Therefore a joint C\,{\sc i}-H$_2$ fit allows us to break the degeneracy and provides significantly tighter constraints on $I_{\rm UV}$ and $n_{\rm H}$. In this work we use the constraint on $I_{\rm UV}-n_{\rm H}$  plane obtained with H$_2$ to set a prior distribution on parameters, when we will analyse excitation of C\,{\sc i} by the CMB radiation. The analysis  is presented in Section\,\ref{sect:ci}

It is believed that CO molecules trace more deep and shielded regions of molecular clouds than  C\,{\sc i}, therefore it was found in a few absorption systems where both H$_2$ and C\,{\sc i} were also detected (see e.g. Noterdaeme et al. 2018b for high redshift DLAs and Burgh et al. 2007, Sheffer et al. 2008 for observations in the Milky-Way). The CO rotational levels are populated by far-infrared radiation (mainly the CMB radiation) and collisions. To take into account collisional pumping we use a constraints on $n_{\rm H}$ and $T_{\rm kin}$ from a joint C\,{\sc i}-H$_2$ fit. 
The analysis of CO excitation is presented in Section\,\ref{sect:co}.   

\section{Excitation of C\,{\sc i} fine-structure levels}
\label{sect:ci}

We use standard assumptions to calculate excitation of C\,{\sc i} fine-structure levels: 
(i) we consider collisions between C\,{\sc i} atoms and H, H$_2$, and He and neglect collisions with electrons. The collisional rate coefficients were taken from Schroder et al. (1991), Staemmler and Flower (1991), and Abrahamsson et al. (2005). 
(ii) We neglect the self-shielding of UV radiation in C\,{\sc i} lines, which are usually optically thin. Radiative pumping rate coefficients were calculated according to the Eq.\,(5) in Silva and Viegas (2002).
(iii) We assume that in diffuse clouds at high redshift  the background radiation at far-infrared wavelength range is determined by the CMB radiation, and excitation of C\,{\sc i} by emission from dust is negligible. We also do not taking into account photon trapping in C\,{\sc i} lines.
(iv) We assume the medium to be homogeneous, i.e. the calculation is performed using a one-zone model: the molecular fraction, species number densities and temperature are constants throughout the cloud.
Therefore, we have 5 fitting parameters: the gas number density, kinetic temperature, UV radiation intensity, CMB temperature, and population of the ground C\,{\sc i} level (${\rm J = 0}$). 
3 of them (number density, kinetic temperature, UV radiation intensity) we constrain by fitting  excitation of H$_2$ rotational levels. For $\log N({\rm H_2})>18$ absorption lines of H$_2$ at J=0,1,2 levels are saturated, and populations of J=0 and J=2 levels are determined mainly by collisions, that gives the excitation temperature $T_{\rm 02}({\rm H_2})\simeq T_{\rm kin}$ (see e.g. Draine 2011). As well we use prior distributions on the number density and UV radiation intensity from a fit to observed populations of H$_2$ J=0,1,2 levels with PDR Meudon models. To estimate parameter values and their statistical uncertainties we use the Markov chain Monte Carlo method.

\begin{figure}
\begin{center}
        \includegraphics[width=0.5\textwidth]{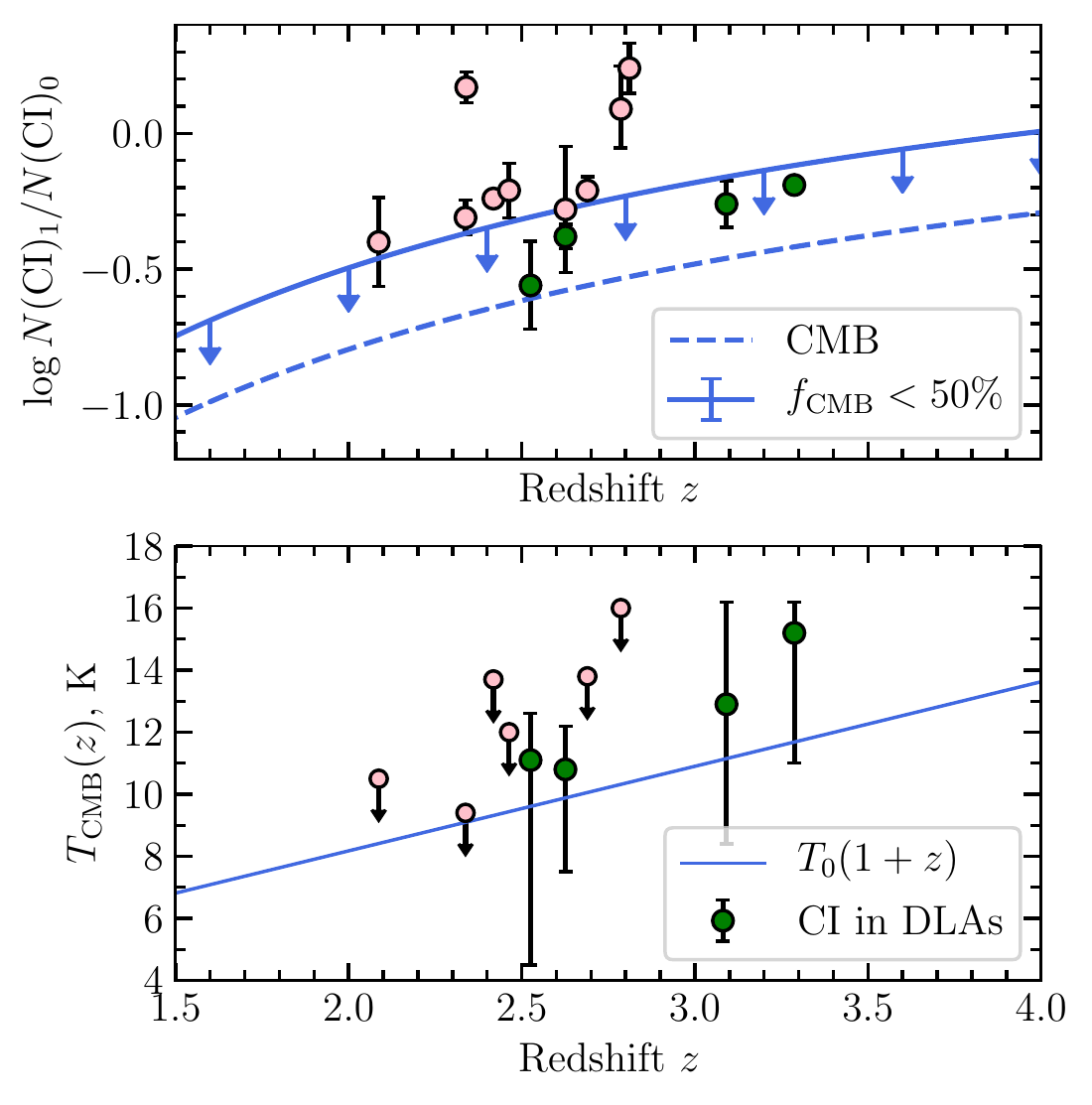}
        \caption{\rm
        Top panel: measurements of relative population of C\,{\sc i}(J=1) level in DLAs as a function of redshift. Green and pink circles represent systems, where contribution of excitation from CMB radiation is higher and lower than 50\% of the total population.
        Blue dashed line represents value of excitation produced by CMB radiation alone. Blue curve with upper limits indicates values of C\,{\sc i}(J=1) population, where CMB contribution equals to 50\%.
        Bottom panel: measurements of the CMB temperature in C\,{\sc i} absorption systems as a function of redshift. Blue line represent the predictions of the standard model $T_{\rm CMB}(z) =2.725{\rm\,K}(1 +z)$.
        }
        \label{fig:ci_corr}
\end{center}
\end{figure}

The results are presented in Table\,\ref{tab:ci}. We measure $T_{\rm CMB}$ in 4 systems, 
where a contribution of the CMB radiation to population of C\,{\sc i}(J=1) level ($f^{\rm CI}_{\rm CMB}$) is significant and consist of about or higher than 50\%. In Fig.\,\ref{fig:ci_corr} we show a calculated contribution of CMB radiation as a function of redshift $z$ (top panel) and compare it with observed relative populations of C\,{\sc i}(J=1) in our sample. The calculation is performed for a cloud with typical physical conditions in diffuse gas: $n_{\rm H}=100\,{\rm cm^{-2}}$, $T_{\rm kin}=100\,K$ and $I_{\rm UV}=1$\,units of Draine field. $T_{\rm CMB}$ is assumed here to be $T_0(1+z)$. 
Blue curve with upper limits shows a border under which the CMB contribution is $>50$\%. There are 4 C\,{\sc i} systems inside this region and they are marked by green color. In other 6 systems the population of C\,{\sc i}(J=1) level is determined mainly by collisions and UV pumping. For them we can set upper limits on $T_{\rm CMB}(z)$. In 3 C\,{\sc i} systems $T_{\rm CMB}(z)$ is not derived, best fit value corresponds to the excitation temperature of \CI. The results are shown in the bottom panel of  Fig.\,\ref{fig:ci_corr}. 
Our estimates of $T_{\rm CMB}(z)$ systematically increase (within 1$\sigma$) the prediction of the standard model $T_0(1+z)$. This may be account to neglecting of excitation from dust emission and photon trapping effect.  




\section{Excitation of CO rotational levels}
\label{sect:co}
In absorption systems at high redshift $z\sim2$ the direct excitation rate of lower rotational levels of CO molecule by CMB photons is much higher than the collisional pumping rate. Therefore, the excitation temperature $T_{\rm exc}({\rm CO})$ was found to be close to the CMB temperature $T_{\rm CMB}$ (Srianand et al. 2008) and it was used to measure the dependence of $T_{\rm CMB}$ on $z$ (Noterdaeme et al. 2011). However, Sobolev et al. (2015) found that  collisional excitation of lower CO rotational levels leads to a small but systematic increase in the $T_{\rm exc}({\rm CO})$ compared to the local $T_{\rm CMB}(z)$ value. In this work we reanalyse rotational excitation of CO in known high-redshift systems taking into account the correction for collisional excitation.

Our model assumes a uniform distribution of molecules over the cloud and takes into account 2 excitation mechanisms: collisions with H, H$_2$, and He and direct excitation by CMB photons. The CO collisional rate coefficients were taken from Walker et al. (2015), Yang et al. (2016), and Cecchi-Pestellini et al. (2002). We have 4 fitting parameters: the gas number density,  kinetic temperature, CMB temperature and population of CO ground level (${\rm J = 0}$). 
We use the Markov chain Monte Carlo method, in which the estimates of $n_{\rm H}$ and $T_{\rm kin}$ from Klimenko and Balashev (2020) are taken into account as a priori distributions.
In the J\,0857$+$1855, J\,1047$+$2057, and J\,1705$+$3543 spectra H$_2$ absorption systems have redshift $z<2$, therefore absorption lines are redshifted outside the observable wavelength range: below the blue end of the UVES spectrograph ($3300$\AA). To estimate the physical conditions in these systems, we analyse the C\,{\sc i} level populations, assumed the kinetic temperature equals to 100\,K. The results are presented in Table\,\ref{tab:co}.

\begin{figure}
\begin{center}
        \includegraphics[width=0.45\textwidth]{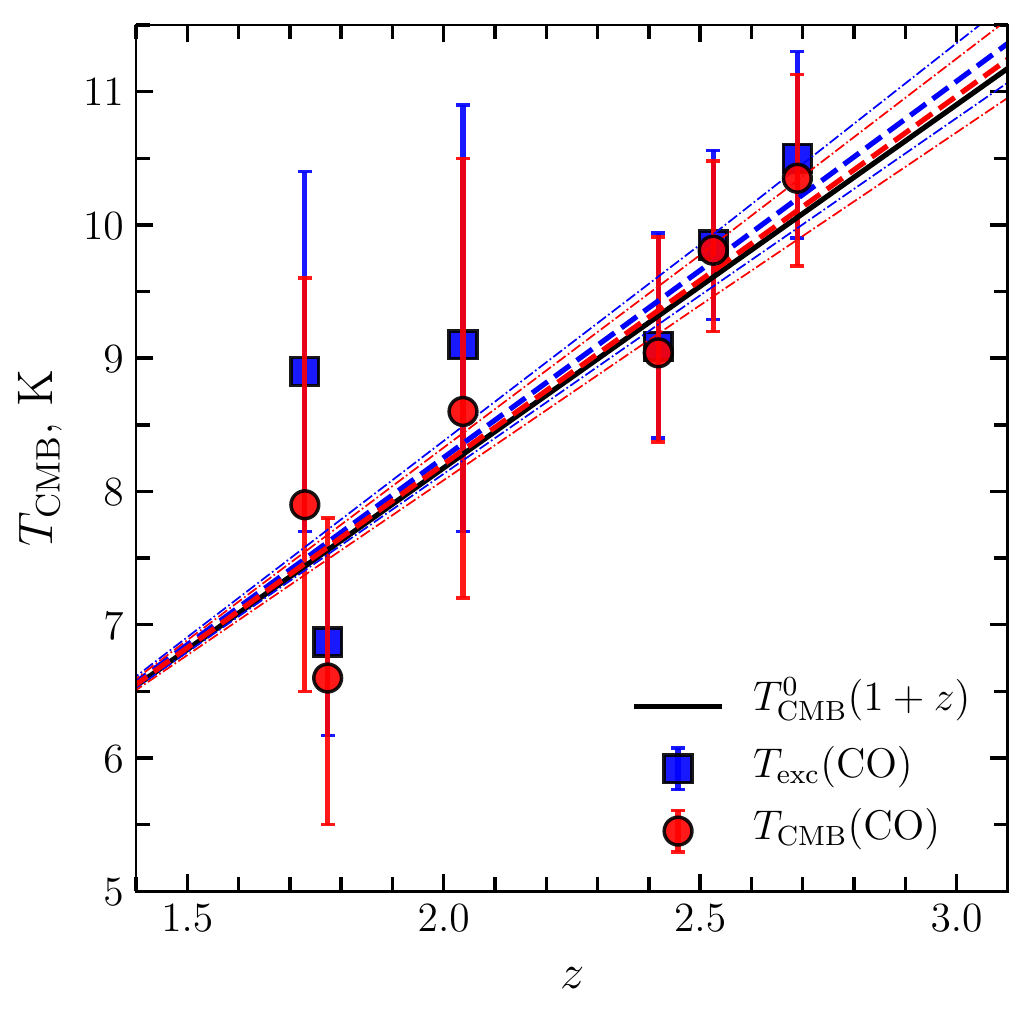}
        \caption{\rm
        Comparison of the CO excitation temperature ($T_{\rm exc}({\rm CO})$, blue squares) and our CMB temperature estimate ($T_{\rm CMB}({\rm CO})$, red circles) made by analyzing the CO rotational level populations with a correction for the collisional excitation. The black solid line indicates the linear relation according to the standard cosmological $\Lambda$CDM model. The blue and red dashed lines indicate the dependence $T_{\rm CMB}(z)=T_0(1+z)^{1-\beta}$ for alternative cosmological models obtained by analyzing the measurements of $T_{\rm exc}({\rm CO})$, $\beta=-0.019\pm0.028$ (blue curve) and $T_{\rm CMB}({\rm CO})$, $\beta=-0.007\pm0.030$  (red curve).
        }
        \label{fig:co_corr}
\end{center}
\end{figure}

In Fig.\ref{fig:co_corr} we compare the CO excitation temperature and the CMB temperature estimated with our method (when we take into account collisional excitation of CO rotational levels). The difference between $T_{\rm CMB}({\rm CO})$ and $T_{\rm exc}({\rm CO})$ turns out to be significant and is $\sim$10\% for J\,0857$+$1855 and J\,1705$+$3543. Given the correction, the new values of  $T_{\rm CMB}({\rm CO})$ show better agreement with the prediction of the standard model.

\section{Results}

\begin{figure*}
\begin{center}
        \includegraphics[width=0.9\textwidth]{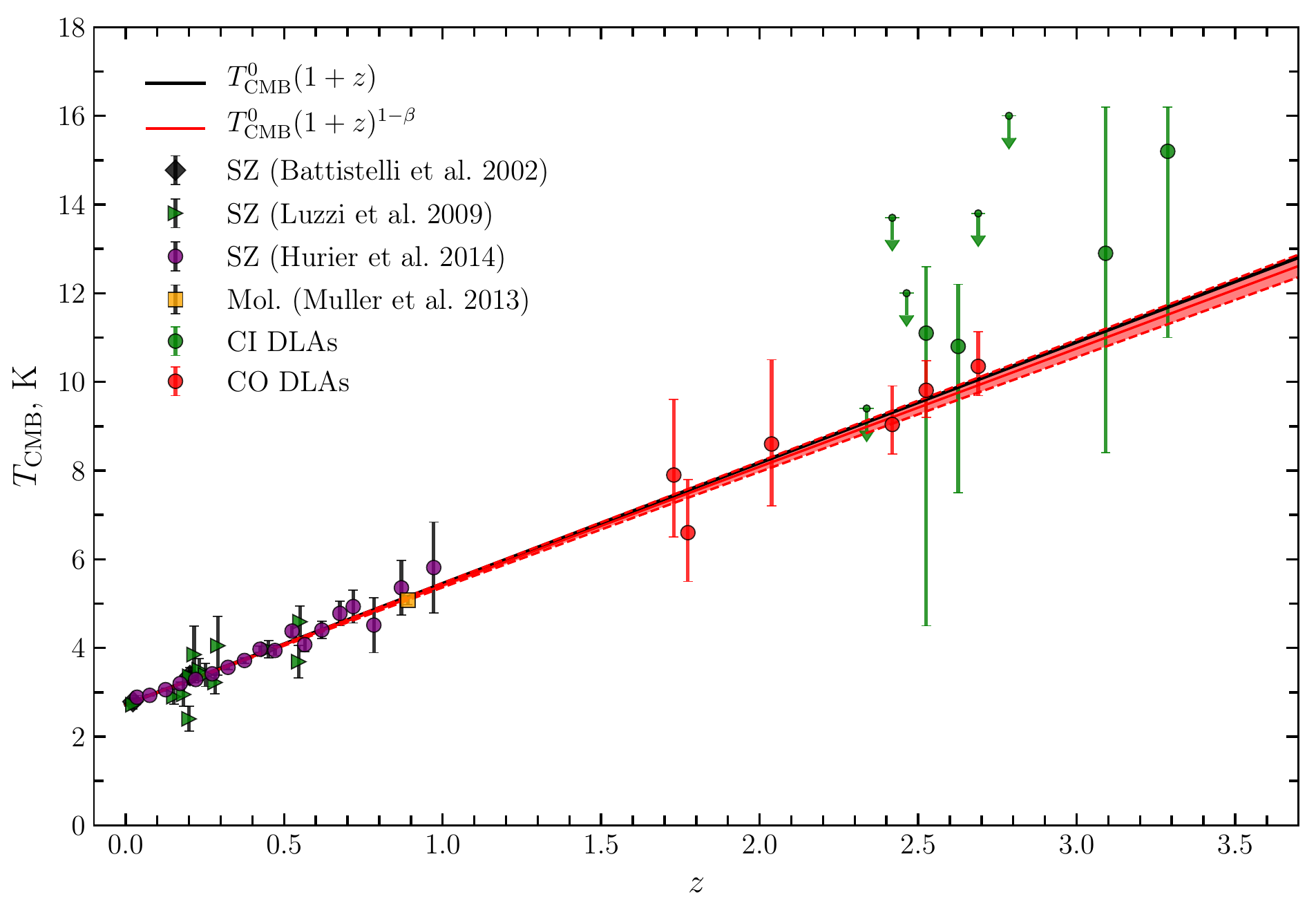}
        \caption{\rm 
        Measurements of the CMB temperature as a function of redshift. The green triangles, violet circles, and black diamonds indicate the measurements based on the analysis of the Sunyaev–Zeldovich effect for galaxy clusters (Luzzi et al. 2009; Hurier et al. 2014; Battistelli et al. 2002). The orange square represents the measurements using molecules in a galaxy at $z=0.89$ (Muller et al. 2013). The green and red circles represent the measurements in the C\,{\sc i} and CO absorption systems in the spectra of quasars (this paper). The black solid line indicates the change in CMB temperature according to the standard $\Lambda$CDM model. The red dashed line and the shaded region indicate the change in $T_{\rm CMB}(z)$  with $z$ for an alternative cosmological model $T_0(1+z)^{1-\beta}$ for the parameter $\beta=0.010\pm0.013$.
        }
        \label{fig:result}
\end{center}
\end{figure*}

The measurements of $T_{\rm CMB}$ derived in high-redshift C\,{\sc i} and CO absorption systems are presented in Figure\,\ref{fig:result} together with other measurements from the analysis of the Sunyaev–Zeldovich effect for galaxy clusters (Battistelli et al. 2002; Luzzi et al. 2009; Hurier et al. 2014) and molecules in a lensing galaxy at $z = 0.89$ in the spectrum of the quasar PKS\,1830$-$211 (Muller et al. 2013). 


We note that the measurements of $T_{\rm CMB}({\rm z})$ derived in the CO and C\,{\sc i} absorption systems 
agree well with the prediction of the standard cosmological model. An accuracy of estimates derived from the rotational excitation of CO is 2-3 times higher than ones based on an analysis of C\,{\sc i} fine-structure populations (see Table\,\ref{tab:result}) . 


Using the expression $T_{\rm CMB}(z) =T_{\rm 0}(1 + z)^{1-\beta}$ (with $T_{\rm 0} = 2.7255 \pm 0.0006$, Fixsen 2009), we determine the parameter $\beta$ for different data samples. The results are presented in Table\,\ref{tab:beta}. The measurements using C\,{\sc i} and CO at high redshift are consistent with the standard relation within the measurement uncertainty, 
$\beta =-0.015^{+0.030}_{-0.028}$. The measurements of $T_{\rm CMB}(z)$ using galaxy clusters give a positive value of $\beta =0.013\pm0.017$. Final estimate using the measurements by the two methods gives a constraint $\beta = 0.010\pm0.013$. 

Our estimate is slightly higher than the estimate, $\beta= 0.006 \pm 0.013$, from Hurier et al. (2014) due to a difference in measurements of $T_{\rm CMB}$ in CO systems. The systematic effect associated with taking into account the collisional excitation of CO is comparable with the statistical uncertainty in the $\beta$ estimate: $\beta = -0.007^{+0.030}_{-0.031}$ versus $\beta= -0.019^{+0.028}_{-0.029}$ for the CO data with and without the collisional correction. 

\begin{table}
\begin{center}
\caption{Measurements of the CMB temperature $T_{\rm CMB}$ in high-redshift C\,{\sc i} and CO absorption systems.}
\label{tab:result}
\begin{tabular}{c|c|c|c}
\hline
QSO & $z_{\rm abs}$ &  $T_{\rm CMB}$(C\,{\sc i})&  $T_{\rm CMB}$(CO) \\
 &  &  K & K \\
\hline 
J\,0857$+$1855 & 1.7293 &  & $9.8^{+0.7}_{-0.6}$  \\
J\,1047$+$2057 & 1.7738 & & $6.6^{+1.2}_{-1.1}$  \\
J\,1705$+$3543 & 2.0377 &&  $8.6^{+1.9}_{-1.4}$ \\  
J\,1232$+$0815 & 2.3377 &  $<9.4$ &  \\
J\,1439$+$1117 & 2.4184 & $<13.7$
& $9.04^{+0.9}_{-0.7}$  \\
J\,1513$+$0352 & 2.4636 &  $<12$\\
J\,0000$+$0048 & 2.5255 &  $11.1^{+1.5}_{-6.6}$ & $9.8^{+0.7}_{-0.6}$  \\
J\,0812$+$3208  & 2.6263 &  $10.8^{+1.4}_{-3.3}$ &  \\
J\,1237$+$0647 & 2.6896 & $<13.8$ & $10.4^{+0.7}_{-0.7}$  \\
J\,0843$+$0221 & 2.7866 &   $<16$ &  \\
J\,2100$-$0641 & 3.0915 & $12.9^{+3.3}_{-4.5}$ & \\
J\,0816$+$1446 & 3.2874 &  $15.2^{+1.0}_{-4.2}$ & \\
\end{tabular}
\end{center}
\end{table}

\begin{table}
\begin{center}
\caption{
Constraints on the parameter $\beta$ characterizing the deviation of the CMB temperature from the standard relation of evolution of the temperature $T_{\rm CMB}=(2.7255\pm0.0006)\times(1+z)^{1-\beta}$. References: [a] - this paper, [b] - Battistelli et al. (2002), [c] - Luzzi et al. (2009), [d] - Hurier et al. (2014), [e] - Müller et al. (2013)
}
\label{tab:beta}
\begin{tabular}{c|c|c}
\hline
Data &  $\beta$ & Reference\\
\hline 
CO & $-0.007^{+0.030}_{-0.031}$ & [a]\\ 
C\,{\sc i} & $-0.077^{+0.130}_{-0.075}$& [a]\\
C\,{\sc i} + CO &$-0.015^{+0.030}_{-0.028}$& [a]\\
SZ & $0.013^{+0.017}_{-0.017}$  & [b, c, d]\\ 
Mol & $0.023^{+0.031}_{-0.032}$ & [e]\\
Mol+ SZ + C\,{\sc i}  & $0.014^{+0.014}_{-0.015}$ & [a, b, c, d, e]\\ 
Mol+ SZ+CO  &$0.011^{+0.014}_{-0.013}$  & [a, b, c, d, e]\\
\hline
Mol + SZ + C\,{\sc i} +CO  &$0.010^{+0.013}_{-0.013}$  & [a, b, c, d, e]\\
\end{tabular}
\end{center}
\end{table} 

\subsection{An independent estimate of $T_{\rm CMB}$ at $z=0$}

By extrapolating the dependence $T_{\rm CMB}({\rm z})$ to zero redshift (z = 0), i.e., by considering $T_{\rm 0}$ as a free parameter when fitting the data with the standard law $T({\rm z}) =T_{\rm 0}(1 + z)$, we obtained an independent estimate of the CMB temperature at the present epoch: $T_{\rm 0} = 2.719\pm0.009$\,K. It agrees well with the currently most accurate estimate obtained from satellite observations, $T_{\rm 0} = 2.7255\pm0.0006$\,K (Fixsen 2009). The results are presented in Table\,\ref{tab:t0}.

\begin{table}
\begin{center}
\caption{
    Measurements of the CMB temperature at present epoch (z = 0) using COBE/FIRAS data (Fixsen 1996, 2009) and the measurements of $T_{\rm CMB}(z)$ in galaxy clusters and absorption systems in the spectra of quasars assuming the standard evolution of $T_{\rm CMB}(z)$.
}
\label{tab:t0}
\begin{tabular}{c|c|c}
\hline 
Data & $T_{\rm CMB}(z=0), K$ &$\beta$ \\
\hline 
Fixsen\,1996  & $2.728~\pm0.004$ & \\
Fixsen\,2009 & $2.7255\pm0.0006$ & \\
CO & $2.75\pm0.11$ & 0 \\
Mol + SZ + C\,{\sc i} +CO & $2.719~\pm0.009$ & 0 \\
\hline \hline
Mol + SZ + C\,{\sc i} +CO  & $2.7255\pm0.0006$ & $0.010\pm0.013$ \\
+ Fixsen\,2009 & & \\
\end{tabular}
\end{center}
\end{table} 

\subsection{A constraint on the equation of state for dark energy}

Considering  alternative cosmological models, the deviation of the redshift dependence of the CMB temperature from the linear relation can be caused by the formation and destruction of photons during the decay of dark energy (see, e.g., Lima et al. 2011). Jetzer et al. (2011) applied this model to describe the CMB temperature measurements and estimated the parameter of the effective equation of state of decaying dark energy ($p = \omega_{\rm eff}\rho$) $\omega_{\rm eff} = -0.97\pm0.03$. Using Eq.\,(22) from Jetzer et al. (2011) and new  measurements in galaxy clusters and absorption systems, we get $\omega_{\rm eff} = -0.991^{+0.014}_{-0.012}$. We assumed that the adiabatic index $\gamma= 4/3$, $T_{\rm 0} = 2.72548\pm0.00057$ (Fixsen 2009) and $\Omega_{\rm m0} = 0.315 \pm 0.007$ (Planck Collaboration 2020). Our estimate is consistent with other estimates, $\omega_{\rm eff} = -0.996\pm0.025$ (Noterdaeme et al. 2011) and $\omega_{\rm eff} = -0.995\pm0.011$ (Hurier et al. 2014).

\section{Conclusion}
\label{sect:concl}
We present measurements of the CMB temperature based on excitation of fine-structure levels of C\,{\sc i} and rotational levels of CO in  high-redshift DLAs. 

In our model of CO excitation we take into account a contribution of collisions, which systematically increases the excitation temperature of J=1 and J=2 CO rotational levels. In diffuse clouds at high redshift $z\sim2$ with thermal pressure  $n_{\rm H}T_{\rm kin}\sim10^4{\rm cm^{-3}K}$  this effect is about 1\,K (or 10\% of $T_{\rm ex}$).
We apply this method to measure of $T_{\rm CMB}$ in 6 CO absorption systems at $z=1.7-2.7$. New measurements of $T_{\rm CMB}$ is in better agreement with the standard cosmological model than previous estimates. 
Analysis of excitation of C\,{\sc i} levels requires independent constraints on the number density and intensity of UV field. We use constraints from an analysis of excitation of low H$_2$ rotational levels, which is nearly orthogonal to the region constrained using C\,{\sc i} and allows to break the degeneracy and provides significantly tighter constraints on these parameters.   
We analysed 12 high redshift DLAs containing both H$_2$ and C\,{\sc i} absorptions and estimate the CMB temperature in 4 systems, where a contribution of CMB radiation to excitation of C\,{\sc i} levels increases 50\%. For other 8 systems we set upper limits on $T_{\rm CMB}$.

Combining our measurements of $T_{\rm  CMB}(z)$ in absorption systems with the constraints from galaxy clusters, we obtained a constraint on the parameter $\beta = +0.010\pm0.013$. This estimate provides a stringent constraint on the parameter of the effective equation of state for decaying dark matter ($\omega_{\rm eff} = -0.991^{+0.014}_{-0.012}$). The CMB temperature at the present epoch was independently estimated by assuming the standard relation $T(z) = T_{\rm 0}(1+z)$: $T_{\rm 0} = 2.719\pm0.009$\,K.

\section{ACKNOWLEDGMENTS}

This work was supported by the Russian Science Foundation (project no. 18-12-00301).

\section*{References}

H. Abgrall, E. Roueff, and Y. Viala, Astron. Astrophys. Suppl. Ser. 50, 505 (1982).\\
E. Abrahamsson, R. V. Krems, and A. Dalgarno, Astrophys. J. 654, 1172 (2007).\\
R. A. Alpher and R. C. Herman, Phys. Rev. 74, 1737 (1948).\\
A. Avgoustidis, R. T. Génova-Santos, G. Luzzi, et al., Phys. Rev. D 93, 043521 (2016).\\
J. N. Bahcall and R. A. Wolf, Astrophys. J. 152, 701 (1968).\\
S. A. Balashev and P. Noterdaeme, Mon. Not. R. Astron. Soc. 478, 7 (2018).\\
S. A. Balashev, A. V. Ivanchik, and D. A. Varshalovich, Astron. Lett. 36, 761 (2010).\\
S. A. Balashev, P. Petitjean, A. V. Ivanchik, et al., Mon. Not. R. Astron. Soc. 418, 357 (2011).\\
S. A. Balashev, P. Noterdaeme, H. Rahmani, et al., Mon. Not. R. Astron. Soc. 470, 2890 (2017).\\
S. A. Balashev, V. V. Klimenko, P. Noterdaeme, et al., Mon. Not. R. Astron. Soc. 490, 2668 (2019).\\
E. S. Battistelli, M. de Petris, L. Lamagna, et al., Astrophys. J. 580, L101 (2002).\\
C. Cecchi-Pestellini, E. Bodo, and N. Balakrishnan, Astrophys. J. 571, 1015 (2002).\\
H. Dekker, S. D’Odorico, A. Kaufer, et al., Proc. SPIE 4008, 534 (2000).\\
Draine B. T., Physics of the Interstellar and Intergalactic Medium. Princeton University Press (2011)\\
R. Fabbri, F. Melchiorri, and V. Natale, Astrophys. Space Sci. 59, 223 (1968).\\
D. J. Fixsen, Astrophys. J. 707, 916 (2009).\\
D. J. Fixsen, E. S. Cheng, J. M. Gales, et al., Astrophys. J. 473, 576 (1996).\\
K. Freese, F. C. Adams, J. A. Frieman, et al., Nucl. Phys. B 287, 797 (1987).\\
G. Gamow, Phys. Rev. 70, 572 (1946).\\
D. S. Gorbunov and V. A. Rubakov, Introduction to the Theory of the Early Universe. Hot Big Bang Theory (World Scientific, Singapore, 2011).\\
R. Guimarges, P. Noterdaeme, P. Petitjean, et al., Astrophys. J. 143, 147 (2012).\\
G. Hurier, N. Aghanim, M. Douspis, et al., Astron. Astrophys. 561, 12 (2014).\\
P. Jetzer, D. Puy, M. Signore, et al., Gen. Relat. Gravit. 43, 1083 (2011).\\
R. A. Jorgenson, A. M. Wolfe, and J. X. Prochaska, Astrophys. J. 722, 460 (2010).\\
V. V. Klimenko and S. A. Balashev, Mon. Not. R. Astron. Soc. 498, 1531 (2020).\\
V. V. Klimenko, S. A. Balashev, A. V. Ivanchik, et al., Mon. Not. R. Astron. Soc. 448, 280 (2015).\\
V. V. Klimenko, P. Petitjean, and A. V. Ivanchik, Mon. Not. R. Astron. Soc. 493, 5743 (2020).\\
S. Ledoux, P. Petitjean, and R. Srianand, Mon. Not. R. Astron. Soc. 346, 209 (2003).\\
J. A. S. Lima, A. I. Silva, and S. M. Viegas, Mon. Not. R. Astron. Soc. 312, 747 (2000).\\
G. Luzzi, M. Shimon, L. Lamagna, et al., Astrophys. J. 705, 1122 (2009).\\
G. Luzzi, D. Génova-Santos, C. J. A. P. Martins, et al., J. Cosmol. Astropart. Phys. 09, 011 (2015).\\
J. Matyjasek, Phys. Rev. D 51, 4154 (1995).\\
A. McKellar, Publ. Astron. Soc. Pacif. 52, 187 (1940).\\
Muller, S., Beelen, A., Black, J. H., et al., Astron. Astrophys. 551, AA109 (2013)\\
P. Noterdaeme, P. Petitjean, C. Ledoux, et al., Astron. Astrophys. 523, 17 (2010).\\
P. Noterdaeme, P. Petitjean, R. Srianand, et al., Astron. Astrophys. 526, L7 (2011).\\
P. Noterdaeme, R. Srianand, H. Rahmani, et al., Astron. Astrophys. 577, 24 (2015).\\
P. Noterdaeme, J. K. Krogager, S. A. Balashev, et al., Astron. Astrophys. 597, 82 (2018).\\
P. Noterdaeme, C. Ledoux, S. Zou, P. Petitjean  et al., Astron. Astrophys. 612, 58 (2018b).\\
A. A. Penzias and R. W. Wilson, Astrophys. J. 142, 419 (1965).\\
F. le Petit, C. Nehme, J. le Bourlot, and E. Roueff, Astrophys. J. Suppl. Ser. 164, 506 (2016).\\ 
Planck Collab. et al., Astron. Astrophys. 641, A6 (2020).\\
A. Ranjan, P. Noterdaeme, J. K. Krogager, et al., Astron. Astrophys. 618, 184 (2018).\\
Y. Rephaeli, Astrophys. J. 241, 858 (1980).\\
K. C. Roth and D. M. Meyer, Astrophys. J. 413, L67 (1993).\\
K. Schroder, V. Staemmler, M. D. Smith, et al., J. Phys. B: At. Mol. Opt. Phys. 24, 2487 (1991).\\
A. I. Silva and S. M. Viegas, Mon. Not. R. Astron. Soc. 329, 135 (2002).\\
A. I. Sobolev, A. V. Ivanchik, D. A. Varshalovich, et al., J. Phys.: Conf. Ser. 661, 012013 (2015).\\
R. Srianand, P. Noterdaeme, C. Ledoux, and P. Petitjean, Astron. Astrophys. 482, L39 (2008).\\
R. Srianand, P. Petitjean, and C. Ledoux, Nature (London, U.K.) 408, 931 (2000).\\
R. Srianand, P. Petitjean, C. Ledoux, et al., Mon. Not. R. Astron. Soc. 362, 549 (2005).\\
V. Staemmler and D. R. Flower, J. Phys. B: At. Mol. Opt. Phys. 24, 2343 (1991).\\
S. S. Vogt, S. L. Allen, B. C. Bigelow, et al., Proc. SPIE 2198, 362 (1994).\\
K. M. Walker, L. Song, B. H. Yang, et al., Astrophys. J. 811, 27 (2015).\\
B. Yang, N. Balakrishnan, P. Zhang, et al., J. Chem. Phys. 145, 034308 (2016).\\
Ya. B. Zeldovich and R. A. Sunyaev, Astrophys. Space Sci. 4, 301 (1969).\\

\end{document}